\documentstyle[twoside,fleqn,espcrc2,babel,english]{article}
\input epsf 
\epsfverbosetrue

\newcommand{\AmS}{{\protect\the\textfont2
  A\kern-.1667em\lower.5ex\hbox{M}\kern-.125emS}}

\title{ Can lattice data for  two heavy-light mesons be understood in
terms of simply two-quark potentials?}

\author{
A.M.~Green\thanks{Presented at Lattice '99 in Pisa, Italy.}, J.~Koponen \address{Department of Physics and Helsinki
Institute of Physics, P.O. Box 9, FIN--00014 University of Helsinki,
Finland
(email: {\tt anthony.green@helsinki.fi}, {\tt jmkopone@rock.helsinki.fi})
}
        and 
 P.~Pennanen\address{Nordita, Blegdamsvej 17, 2100 Copenhagen \O,
Denmark (email: {\tt petrus@hip.fi})
}}
       
\begin{document}

\begin{abstract}
By comparing lattice data for the two heavy-light meson system
 ($Q^2\bar{q}^2$) with a standard many-body approach employing only
interquark potentials, it is shown that the use of {\em unmodified
two-quark} potentials leads to a gross overestimate of the binding
energy.
\end{abstract}

\maketitle

\section{INTRODUCTION}
 
In this conference the emphasis is on the study of {\em single} hadrons,
with virtually nothing being said about systems involving more than
three quarks. This is unfortunate, since it avoids much of particle
physics and all of nuclear physics. It is, therefore, desirable to
attempt to bridge this gap. Since the ability to perform accurate
lattice simulations for systems containing many quarks is severely
limited -- with four quarks being the current maximum -- the challenge
is to see how (if) standard many-body techniques can be developed for 
the new situation of interacting quarks and gluons. Here we immediately
meet a cultural difference.
For particle physicists, the use of simply two-quark potentials in multiquark
problems is not a discussion point, since they know (believe)
that it is not possible. In spite of this, there are many-body physicists,
who still believe (hope) that this is not so. 

\begin{figure}[htb]
\makebox[55mm]{\rule[-21mm]{0mm}{43mm}}
\includegraphics{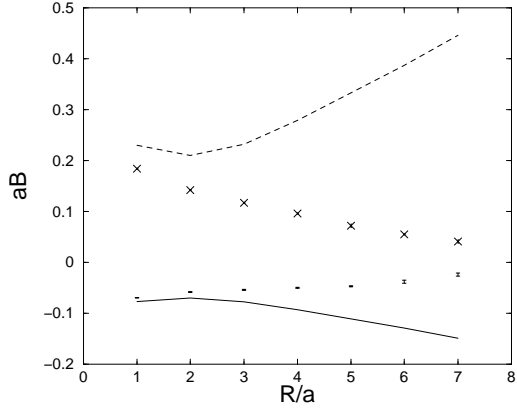}
\caption{The case of four static quarks $(Q^4)$ in SU(2)
placed at the corners of
squares of side $R/a$, where $a\approx 0.12$ fm. This shows the binding
energy in lattice units ($aB$) as a function of $R/a$.  
The lattice results are for $\beta=2.4$
on a $16^3\times 32$ lattice with the dots (crosses) showing the
ground (excited) state energies -- with error bars.
 The two curves are for the model resulting  in Eq. \ref{VN} (with
$k_f=0$) and correspond to the ground (solid line) and excited (dashed line)
states. This figure is similar to the one in ~\protect\cite{gmp}}
\label{fig:1}
\end{figure}

\begin{figure}[htb]
\vspace{9pt}
\makebox[55mm]{\rule[-21mm]{0mm}{43mm}}
\includegraphics{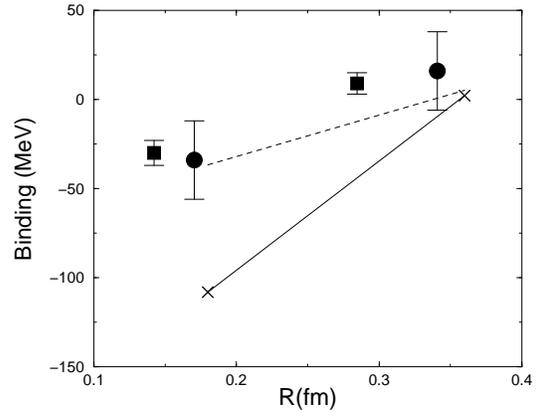}
\caption{Comparison between the spin independent part
$(V_0)$ of the $Q^2\bar{q}^2$ binding energies
calculated on a lattice ~\protect\cite{mp} (solid circles -- quenched
approximation with $a=0.170$ fm)/~\protect\cite{UKQCD}
(solid squares -- with dynamical fermions and $a=0.142$ fm)
and the model in the weak coupling limit ($k_f=0$) -- solid line.
The dashed line shows the model prediction for $k_f=0.25$. These
lines are to guide the eye. The dynamical fermion
data is not used in any fit. It is simply included to show that it is
qualitatively consistent with the quenched data but with considerably
smaller error bars. }
\label{fig:2}
\end{figure}

Over the last few years
attempts have been made to clarify this situation by comparing the
{\em exact} energies of four-quark systems -- as calculated on a lattice
-- with standard many-body models using only two-quark potentials~[1].
The four-quark system was studied for two reasons:\\
i) Accurate lattice simulations are still possible.\\
ii) It is the first step in the description of interacting hadrons, 
since the system can be partitioned into two colour singlets.
The outcome from those pilot comparisons was  
 that --  "beyond all reasonable doubt" -- the resulting
binding energies are grossly overestimated by the models [Figure \ref{fig:1}] and 
that a four-quark
form factor is necessary. 
However, this conclusion was based on lattice
data for four static quarks in quenched SU(2). In the present 
work~\cite{mp,gkp} most of
these approximations  have been removed by using SU(3) with two heavy-light
mesons i.e. $Q^2\bar{q}^2$ compared with the earlier $Q^2\bar{Q}^2$.
Furthermore, the gauge field is now  treated in the unquenched
approximation. But this latter improvement only has a small effect. 
The corresponding many-body model now needs some
additional assumptions before a  comparison with the lattice data is
possible. Perhaps the most serious is the use of a non-relativistic
kinetic energy, even though the light quark mass is approximately that
of the strange quark. However, in the binding energy there is a
considerable cancellation between the four- and two-quark kinetic
energies, which then removes much of the kinetic energy effect from the
binding energy.
 Also the model is developed only for the spin-independent
contribution to the binding energy. This requires an averaging of the
lattice data. Both of these assumptions are not thought to qualitatively
effect the outcome. The main defect of the calculation is that, at
present, comparison can only be made for small distances between the
two static quarks --  the region where lattice data could suffer from a
lack of rotational invariance. However, in spite of these shortcomings,
the result is that the use of simply two-quark potentials again 
overestimates the lattice binding by upto a factor of three -- a result
that is not expected to be qualitatively changed by other models without
multi-quark interactions [Figure \ref{fig:2} -- solid line]. 
Inclusion of a four-quark interaction term can then remove the 
discrepancy [Figure \ref{fig:2} -- dashed line].

The reason why the "data-model" comparison is good for small $R/a$ in
Figure \ref{fig:1} and not in Figure \ref{fig:2} is simply because the 
$Q^2\bar{q}^2$ energy -- even for small $R$ -- is dominated by potential
energy contributions, where the $Q\bar{q}$
and $\bar{q}\bar{q}$ separations are  $\approx 3a$. As seen in 
Figure \ref{fig:1}, at such separations the model already becomes too attractive
in the absence of the form factor $f$.

\section{MODEL}

In Refs.~\cite{gmp,gkp,model1} a model was developed for understanding the lattice
energies of four static quarks \\
$Q({\bf r_1})Q({\bf r_2})\bar{Q}({\bf r_3})\bar{Q}({\bf r_4})$
in terms of two-quark potentials. This model, in its simpliest form, was 
constructed in terms of the two basis states that can be made by
partitioning the four quarks into two color singlets - namely -
\begin{eqnarray}
\label{AB}
A&=&[Q(1)\bar{Q}(3)][Q(2)\bar{Q}(4)] \nonumber \\ 
B&=&[Q(1)\bar{Q}(4)][Q(2)\bar{Q}(3)], 
\end{eqnarray}
where $[...]$ denotes a color singlet.
These two states are not orthogonal and have a normalisation matrix of
the form.

\begin{equation}
\label{Nf}
\bf{N}(f)=\left( \begin{array}{ll}
 1   & \frac{1}{3}f \\
\frac{1}{3}f & 1\\
\end{array} \right).
\end{equation}
In the extreme weak coupling limit the parameter $f=1$ and in the strong
coupling limit $f=0$. However, for intermediate situations 
it is parametrised as
\begin{equation}
\label{fexp}
f({\bf r_1},{\bf r_2},{\bf r_3},{\bf r_4})
=\exp[-b_s k_f 
S({\bf r_1},{\bf r_2},{\bf r_3},{\bf r_4})],
\end{equation}
where $b_s$ is the string energy,
$S(\bf{r_1},\bf{r_2},\bf{r_3},\bf{r_4})$ is an area defined by the
positions of the quarks and $k_f$ is a free parameter. 
A single value for $k_f$ was then
capable of giving a reasonable understanding for 100 pieces of data --
the ground and first excited states of configurations from six different
four-quark geometries calculated on a $16^3\times 32$ lattice.
In this model the interaction between the quarks is expressed as a
potential matrix of the form
\begin{equation}
\label{Vf}
\bf{V}(f)=\left( \begin{array}{ll}
 v(13)+v(24)   & fV_{AB} \\
fV_{AB} & v(14)+v(23) \\
\end{array} \right),
\end{equation}
where
$V_{AB}$ has the form  expected in the weak coupling limit with the
one-gluon-exchange-potential
\begin{equation}
\label{OGE}
V=-\frac{1}{3}\sum_{i\leq j}\lambda_i \lambda_jv(ij) \ \ {\rm and } \ \
v(ij)=-\frac{e}{r_{ij}}.
\end{equation}
Away from the weak coupling limit, $f$ is no longer unity and in addition
$v(ij)$ is taken to be the full two quark potential.
The  energy of the four static quarks -- a function of the $\bf{r_i}$ -- is
then given by diagonalising
\begin{equation}
\label{VN}
|{\bf V}(k_f)-E(4, {\bf r_i} ,k_f){\bf N}(k_f)|\psi=0.
\end{equation} 
This model is easy to generalise to $Q^2\bar{q}^2$ by integrating over
the positions of the two light quarks using a suitable variational 
wavefunction~\cite{gkp}. This results in the need to diagonalise
\begin{equation}
\label{KVN}
|{\bf K}(k_f)+{\bf V}(k_f)-E(4,R, k_f){\bf N}(k_f)|\psi=0,
\end{equation} 
where ${\bf K}(R, k_f)$ is the kinetic energy and $R$ is the distance between
the two static quarks.
In order to carry out the radial integrals in ${\bf K }(R, k_f), \  {\bf V}(R,k_f)$
and ${\bf N}(R, k_f)$, the function 
$f$ in Eq.~\ref{fexp} was now reduced to the more symmetric form
\begin{equation}
\label{fsym}
f=\exp\left[-k_f b_s \sum_{i\leq j}r_{ij}^2 \right].
\end{equation}   
Again $k_f$ is a free parameter, which should be adjusted to fit the
four-quark lattice energies.
However, now only the lattice data for $R=a$ is sufficiently accurate to
extract a value of $k_f$ -- see Figure~\ref{fig:2}.

The present model is now being developed to include:\\ 
1) a semi-relativistic approximation, in which for the two light quarks the
kinetic energy operator in momentum space is simply replaced by 
$\sqrt{p^2_i+m^2_q}-m_q$.\\
2) a spin dependence in the potential between the two light quarks.

This model, although very simple, contains the same basic
assumptions made in the more elaborate many-body models that incorporate
kinetic energy e.g. the Resonating Group Method~\cite{Yaz}.
It is, therefore, reasonable that this simplified model can to some
extent check the validity of its more elaborate counterparts.

\section{CONCLUSION}
The comparison between lattice data and a model for the $Q^2\bar{q}^2$
system  supports (confirms) the earlier result -- with 
four static quarks in quenched SU(2) -- that four-quark energies cannot be
described simply in terms of two-quark potentials and that attempts to
do so could lead to a large overestimate of the binding energy.
One way of overcoming this problem is to include into the model a
four-quark form factor -- as seen in Figure 2.

\end{document}